\documentclass[twocolumn,10pt]{revtex4}

\usepackage{eurosym}
\usepackage{amsfonts}
\usepackage{amsmath}
\usepackage{amssymb,epsf}
\usepackage{color}
\usepackage{graphicx}
\usepackage{epstopdf}
\usepackage{float}
\usepackage{caption}
\usepackage{subfig}

\begin{document}

\title{The shadow of black holes in $F(R)$-ModMax theory with cosmic strings }
\author{Ahmad Al-Badawi}
\email[Email: ]{ahmadbadawi@ahu.edu.jo 
}
\affiliation{Department of Physics, Al-Hussein Bin Talal University, 71111,
Ma'an, Jordan.}

\begin{abstract}
This work explores the shadow of a black hole within the framework of $F(R)$-ModMax gravity coupled with a cloud of strings. The Einstein field equations are solved for a nonlinear ModMax electromagnetic source in the context of $F(R)$ gravity and a string cloud. From this solution, we  obtain analytical expressions for the photon sphere and shadow radii. Our findings reveal that the interplay between nonlinear electrodynamics, $F(R)$ gravity, and the string cloud significantly alters spacetime geometry, leading to distinct dynamical behaviors for test particles while also amplifying the shadow radius. These results underscore the critical role of cosmic strings effects and modified gravity parameters in shaping black hole shadows.

\end{abstract}

\maketitle

\section{Introduction}

In recent decades, the pursuit of comprehending the universe's accelerated expansion, frequently ascribed to dark energy (DE), has motivated the investigation of modified theories of gravity (MOG). The MOG theories extend general relativity (GR) and provide convincing alternatives to the $\Lambda$CDM model based on cosmological constants. The $\Lambda$CDM model \cite{nq9,nq10} provides a simple explanation for DE, also known as the cosmological constant in general relativity. The basic cosmological model successfully explains nearly all of cosmic history. However, it suffers from the  so-called cosmological constant problem, which explains why cosmological constants are so small and on the scale of the critical density of the Universe.   \\ One of the simplest
extensions of the modified gravity theory is $F(R)$ theory of gravity \cite{mxxx1,mxxx2}, where the scalar curvature $R$ is replaced by an arbitrary function of $R$  \cite{mx3,mx4}. Motivations for studying $F(R)$ theory of gravity include explaining the Universe's accelerated expansion and structure formation without invoking DE or dark matter \cite{mx5,mx6}. Additionally, $F(R)$ theory aligns with Newtonian and post-Newtonian approximations \cite{mx7,mx8} and captures key features of higher-order gravity through its action \cite{mx9}.  By introducing some consistent models \cite{NojiriO2003,NojiriO2011}, this modified theory of gravity is able to describe the evolution of the whole universe. Furthermore, $F(R)$ gravity theory can explain various phenomena observed in cosmology and astrophysics  \cite{Mod1,Mod3,Mod5,Mod6,Mod7,Mod7a,Mod7b}.

On the other hand, the Modified Maxwell (ModMax) theory, introduced by Bandos et al.  \cite{mx10}, is a nonlinear generalization of Maxwell electrodynamics that retains conformal invariance. Governed by a dimensionless parameter $\gamma$, this theory smoothly transitions between Maxwell’s linear regime ($\gamma = 0$) and a nonlinear regime while preserving both conformal symmetry and electric-magnetic duality rotations \cite{mx10}. What distinguishes ModMax from other nonlinear extensions is its unique determination by these symmetries, providing a robust framework for investigating conformally invariant phenomena beyond classical electrodynamics \cite{mx10}. A number of aspects of ModMax electrodynamics and the corresponding black hole (BH) solutions have been extensively investigated \cite{new11,md8,mxx26,mxx10}. Recently, an exact analytical solution for BHs is obtained by coupling ModMax nonlinear electrodynamics and $F(R)$ gravity \cite{mxx16}. 

Cosmic strings, arising from symmetry-breaking phase transitions in the early universe, are topological defects with potential implications for large-scale structure formation \cite{ TK1,AV2}. They may have seeded primordial density inhomogeneities, influencing galaxy and cluster formation \cite{DM1}. Letelier’s cloud of strings (CS) model (a pressure-less perfect fluid) \cite{mbhh1} has been widely used to derive BH solutions in GR and extended gravity theories (e.g., Einstein-Gauss-Bonnet, Lovelock gravity) \cite{mbhh1, mbhh2, PSL3,DVS2, FFN1, FFN2, MC1,NDSS}. These solutions generalize classical results like the Schwarzschild metric. The CS framework provides a robust alternative to point-particle models, aligning with string theory’s proposition that 1-dimensional strings could be fundamental cosmic constituents. Recent studies highlight the gravitational role of cosmic and fundamental strings, emphasizing their astrophysical and cosmological significance \cite{JLS}.

  This paper introduces a novel metric for describing a BH in $F(R)$-ModMax theory with CS. As a result of this metric function, we are able to study the combined impact of $F(R)$-ModMax theory and CS effects on the theory of BHs.  The motivation for this research derives from the realization that true astrophysical BHs are unlikely to exist in isolation, but rather immersed in complex surroundings comprising diverse types of matter and energy. While prior research has focused on BHs using either $F(R)$-ModMax theory or CS, the combined system represents a more realistic and physiologically justified scenario that encompasses the complexities of actual astrophysical environments. {Furthermore, the combination of ModMax with modified gravity theory $F(R)$ and CS defect addresses multiple fundamental issues simultaneously: to properly understand the modified gravity and explore its validity and the singularity problem through the length scale parameter $\alpha$ provides a unified framework for testing both quantum gravity signatures and 
modified spacetime geometry in observational contexts. This composite approach offers richer phenomenology than individual components alone, potentially enabling discrimination between different theoretical models through distinct observational signatures in shadow measurements and gravitational wave ringdown phases. An extensive body of literature exists on shadow analysis; for relevant models, we refer to \cite{new1,new2,new3}}

  The outline of this work is as follows: In Sec.~\ref{sec2} we introduce the Einstein field equations 
 by coupling ModMax NLED and $F(R)$ gravity  with CS and derived  the static spherical symmetric BH solution. In  Sec.~\ref{sec3}, we investigate  the BH shadow and obtain analytic expressions for both photon sphere and shadow radius.  Conclusions are discussed in Section~\ref{sec5}. 

\section{BHs in $F(R)$-ModMax theory with Cosmic strings} \label{sec2}

In this section, we introduce the coupling of the ModMax field with $F(R)$ gravity and CS. In four-dimensional spacetime, the
action of $F(R)$-ModMax theory with CS is given by 
\begin{equation}
S=\frac{1}{16\pi }\int_{\partial \mathcal{M}}d^{4}x\sqrt{-g}%
\left[ F(R)-4(\mathcal{L}_{MM}+\mathcal{L}_{CS})\right] ,  \label{actionF(R)}
\end{equation}%
where $g=det(g_{\mu \nu})$  stands for the determinant of the metric tensor, $F(R)=R+f\left( R\right) $, in which $R$ and $f\left( R\right) $,
respectively, are scalar curvature and a function of scalar curvature. $\mathcal{L}_{MM}$ and $\mathcal{L}_{CS}$ are the Lagrangians of ModMax and CS, respectively.  The ModMax Lagrangian $\mathcal{L}_{MM}$ is defined as \cite{mx10,new11} 
\begin{equation}
\mathcal{L}_{MM}=\frac{1}{2}\left( \mathcal{S}\cosh \gamma -\sqrt{\mathcal{S}^{2}+%
\mathcal{P}^{2}}\sinh \gamma \right) ,  \label{ModMaxL}
\end{equation}%
where $\gamma $  is a dimensionless parameter of the ModMax theory and $\mathcal{S}=\frac{\mathcal{F}}{2}$, and $\mathcal{P}=\frac{\widetilde{%
\mathcal{F}}}{2}$ are,
respectively, a true scalar, and a pseudoscalar. Here  $\mathcal{F}=F_{\mu\nu}F^{\mu\nu}$ is Maxwell's invariant and $F_{\mu\nu}=\partial_\mu A_\nu-\partial_\nu A_\mu$ to be the electromagnetic field, $A_\mu$ being the gauge potential. As a further point, the invariant $\Tilde{\mathcal{F}}$ can be expressed within $\Tilde{\mathcal{F}}=F_{\mu\nu}\Tilde{F}^{\mu\nu}$.

which are defined in the
following forms, where $\mathcal{F}=F_{\mu \nu }F^{\mu \nu }$ is the Maxwell invariant ($F_{\mu
\nu }=\partial _{\mu }A_{\nu }-\partial _{\nu }A_{\mu }$ (where $A_{\mu }$
is the gauge potential) is the electromagnetic tensor). In addition, $%
\widetilde{\mathcal{F}}$ equals to $F_{\mu \nu }\widetilde{F}^{\mu \nu }$,
where $\widetilde{F}^{\mu \nu }=\frac{1}{2}\epsilon _{\mu \nu }^{~~~\rho
\lambda }F_{\rho \lambda }$. 

 To describes strings like objects \cite{mbhh1,mbhh2}, we use the Nambu-Goto action 
\begin{equation}
    S^{CS}=\int \sqrt{-\zeta}\,\mathcal{M}\,d\lambda^0\,d\lambda^1=\int \mathcal{M}\sqrt{-\frac{1}{2}\Sigma^{\mu \nu}\,\Sigma_{\mu\nu}}d\lambda^0\,d\lambda^1,\label{ac1}
\end{equation}
where $\mathcal{M}$ is the dimensionless constant which characterizes the string, ($\lambda^0\,\lambda^1$) are the time
like and spacelike coordinate parameters, respectively \cite{PSL3}. $\zeta$ is the determinant of the induced metric of the strings world sheet given by $\zeta=g^{\mu\nu}\frac{ \partial x^\mu}{\partial \lambda^a}\frac{ \partial x^\nu}{\partial \lambda^b}$. Here $\Sigma^{\mu\nu}=\epsilon^{ab}\frac{ \partial x^\mu}{\partial \lambda^a}\frac{ \partial x^\nu}{\partial \lambda^b}$ is bivector related to string world sheet, where $\epsilon^{ ab}$ is the second rank Levi-Civita tensor which takes the non-zero values as $\epsilon^{ 01} = -\epsilon^{ 10} = 1$.  Moreover, $T_{\mu \nu}=2\,\partial L/\partial g^{\mu\nu}$, and then $\partial_\mu (\sqrt{-g}\rho\,\Sigma^{\mu\nu})=0,$  where $\rho$ is the density, which describes the
 case of a CS \cite{PSL3}, and the $\Sigma^{\mu\nu}$ is the function of radial distance. The non-vanishing component of $\Sigma^{\mu\nu}$ is $\Sigma^{tr}=\Sigma^{rt}$. Hence, the energy-momentum tensor becomes $T^t_t=T^r_r=-\rho\,\Sigma^{tr}$. Thus using $\partial_t(r^2\,\Sigma^{tr})=0$, \begin{equation}
  T^t_t=T^r_r=\frac{\alpha}{r^2}  , 
 \end{equation}
where $\alpha$ is an integration constant related to strings or CS parameter.

To create an electrically charged BH in $F(R)$ solution suitable for the action (\ref{actionF(R)}), we must process only the electric field source of the ModMax theory while imposing the restriction $\mathcal{P}=0$. Varying the action (\ref{actionF(R)}) leads to field equations,
\begin{eqnarray}\label{EqF(R)1}
R_{\mu \nu }\left( 1+f_{R}\right) -\frac{g_{\mu \nu }F(R)}{2}+\left( g_{\mu
\nu }\nabla ^{2}-\nabla _{\mu }\nabla _{\nu }\right) f_{R}\\=8\pi \left(T_{\mu\nu}^{\text{MM} }-T_{\mu\nu}^{CS}\right),   \nonumber 
\end{eqnarray}
\begin{equation}
    \partial _{\mu }\left( \sqrt{-g}\widetilde{E}^{\mu \nu }\right) =0,\label{EqF(R)2}
\end{equation}
where $f_{R}=\frac{df(R)}{dR}$ and $\mathcal{R}_{\mu\nu}$ is the Ricci tensor, $T_{\mu\nu}^{\text{MM}}$ is the energy-momentum tensor for the ModMax theory, and $T_{\mu\nu}^{CS}$ is the energy-momentum tensor for the CS. \\The
energy-momentum tensor of ModMax theory is given by 
\begin{equation}
4\pi\, T_{MM}^{\mu\nu}=\left( F^{\mu \sigma }F_{~~\sigma }^{\nu
}e^{-\gamma }\right) -e^{-\gamma }\mathcal{S}g^{\mu \nu },  \label{eq3}
\end{equation}
The energy-momentum tensors for the CS is given by 
 \begin{equation}
   T_{\mu\nu}^{CS}=2 \frac{\partial}{\partial g_{\mu \nu}}\mathcal{M}\sqrt{-\frac{1}{2}\Sigma^{\mu \nu}\,\Sigma_{\mu\nu}} =\frac{\rho \,\Sigma_{\alpha\nu}\, \,\Sigma_{\mu}^\alpha }{\sqrt{-\gamma}}, 
 \end{equation}
where $\rho$ is the proper density of the CS.
And $\widetilde{E}_{\mu \nu }$ in Eq. (\ref{EqF(R)2}), is defined as 
\begin{equation}
\widetilde{E}_{\mu \nu }=\frac{\partial \mathcal{L}}{\partial F^{\mu \nu }}%
=2\left( \mathcal{L}_{\mathcal{S}}F_{\mu \nu }\right) ,  \label{eq3b}
\end{equation}%
where $\mathcal{L}_{\mathcal{S}}=\frac{\partial \mathcal{L}}{\partial 
\mathcal{S}}$. So, the ModMax field equation (Eq. (\ref{EqF(R)2})) for the
electrically charged case reduces to 
\begin{equation}
\partial _{\mu }\left( \sqrt{-g}e^{-\gamma }F^{\mu \nu }\right) =0.
\label{Maxwell Equation}
\end{equation}
Our objective is to construct BH modeling solutions that are relevant to model (\ref{actionF(R)}) by considering static, spherically symmetric spacetime as 
\begin{equation}
ds^{2}=-g(r)dt^{2}+\frac{dr^{2}}{g(r)}+r^{2}\left( d\theta ^{2}+\sin
^{2}\theta d\varphi ^{2}\right) ,  \label{Metric}
\end{equation}%
in which $g(r)$ defines as the metric function. Let us assume the
constant scalar curvature $R=R_{0}=$ constant, then the trace of the
equation (\ref{EqF(R)1}) turns to 
\begin{equation}
R_{0}\left( 1+f_{R_{0}}\right) -2\left( R_{0}+f(R_{0})\right) =0,
\label{R00}
\end{equation}%
where $f_{R_{0}}=$ $f_{R_{\left\vert _{R=R_{0}}\right. }}$. Solving the
equation (\ref{R00}) in terms of  $R_{0}$ leads to 
\begin{equation}
R_{0}=\frac{2f(R_{0})}{f_{R_{0}}-1}.  \label{R0}
\end{equation}

Replacing Eq. (\ref{R0}) within Eq. (\ref{EqF(R)1}) then the  equations of motion $F(R)$-ModMax
theory's with CS can be found in the following format 
\begin{equation}
R_{\mu \nu }\left( 1+f_{R_{0}}\right) -\frac{g_{\mu \nu }}{4}R_{0}\left(
1+f_{R_{0}}\right) =8\pi \left(T_{\mu\nu}^{\text{MM} }-T_{\mu\nu}^{CS}\right).  \label{F(R)Trace}
\end{equation}

To obtain a radial electric field, we take the following form for the gauge
potential $A_{\mu }=h\left( r\right) \delta _{\mu }^{t}$.

By utilizing the provided gauge potential and equations (\ref{Maxwell
Equation}) and (\ref{Metric}), we found $h(r)=-\frac{q}{r}$, where $q$ represents an integration constant that is associated with the
electric charge.

Finally,  solving the field equations, we can obtain the metric function $g(r)$ as follows: 
\begin{equation}
g(r)=1-\alpha-\frac{m_{0}}{r}-\frac{R_{0}r^{2}}{12}+\frac{q^{2}e^{-\gamma }}{\left(
1+f_{R_{0}}\right) r^{2}},  \label{g(r)F(R)}
\end{equation}%
where $m_{0}$ is an integration constant that is connected to the BH's geometric mass and $\alpha$ is the CS parameter. 
Furthermore, any of the field equations (\ref{F(R)Trace}) is satisfied by
the obtained solution (\ref{g(r)F(R)}). We should limit ourselves to $%
f_{R_{0}}\neq -1$ in order to have physical solutions. \\ It is possible to recover several BH solutions as special cases of the generalized spacetime in  Eq. (\ref{g(r)F(R)}):
\begin{eqnarray*}
   &\alpha=0 \Rightarrow 
 \text{ $F(R)$-ModMax}\, \cite{mxx16}, 
\\& f_{R_0}=0 \Rightarrow  \text{ ModMax with CS}, \,
 \,\\& f_{R_0}=0=\alpha \Rightarrow  \text{ ModMax}\, \cite{mxx10}, \\ &f_{R_0}=0=\alpha=\gamma \Rightarrow \text{  Reissner-Nordstr\"{o}%
m-(A)dS .} 
\end{eqnarray*}

\begin{figure}
    \centering
    \includegraphics[width=0.8\linewidth]{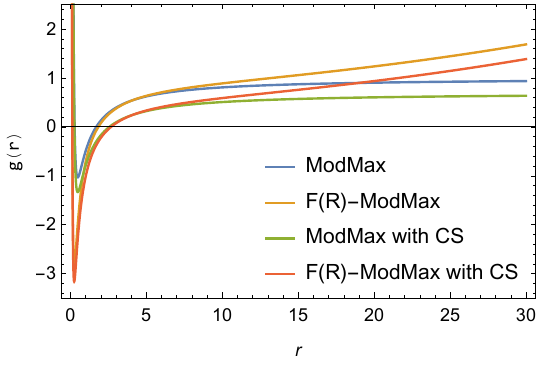}
    \caption{ A comparison of the metric function $g(r)$ for different BHs. Here, we set $m_0=1,\gamma=0.5,f_{R_0}=0.9=q,R_0=-0.01$ and $\alpha=0.3$}
    \label{lapseall}
\end{figure}

 Figure \ref{lapseall} compares the metric functions of various BH solutions: ModMax, $F(R)$-ModMax, ModMax with a CS, and $F(R)$-ModMax with a CS. Notably, all these solutions can exhibit at most two horizons.

\section {Shadow of BH in $F(R)$-ModMax theory with CS } \label{sec3}

To study the shadow cast by the  BHs in $F(R)$-ModMax theory with CS, we need to obtain the equation of the photon sphere. Thus, we consider the  photon sphere equation, namely
\begin{equation}
r_{ph}\,g^{\prime }\left( r_{ph}\right) -2g\left( r_{ph}\right) =0,
\label{psr}
\end{equation}
in terms of photon sphere radius ($r_{ph}$). After we substitute the {metric} function, Eq. (\ref{g(r)F(R)}), we obtain the equation for the photon sphere namely,  
\begin{equation}
2(\alpha-1)(1+f_{R_{0}})r^2+3\,m_0(1+f_{R_0})r-4q^2e^{-\gamma}=0
. \label{eps1}
\end{equation}
The analytical solution of Eq. (\ref{eps1}) is:
 \begin{eqnarray}
 r_{ph}=\frac{-3\,m_0-\sqrt{9\,m_0^2+\frac{32\, q^2(\alpha-1)e^{-\gamma}}{1+f_{R_0}}}}{4(\alpha-1)}. \label{bb16c}
\end{eqnarray}
 When switching off the parameter  $\alpha=0$, then  Eq. (\ref{bb16c}) reduces to the photon sphere of $F(R)$-ModMax theory. In the limit of $f_{R_0}=0=\alpha$ we obtain the photon sphere in the case of ModMax BH.\\  
Our shadow radius $R_s$ can be directly calculated from the orbit radii of the photons
as  \begin{equation}
    R_s=\frac{r_{ph}}{\sqrt{1-\alpha-\frac{m_{0}}{r_{ph}}-\frac{R_{0}r^{2}_{ph}}{12}+\frac{q^{2}e^{-\gamma }}{\left(
1+f_{R_{0}}\right) r^{2}_{ph}}}} .\label{shadeq1}
\end{equation}

\begin{figure}[ht!]
    \centering
    {\centering{}\includegraphics[width=0.85\linewidth]{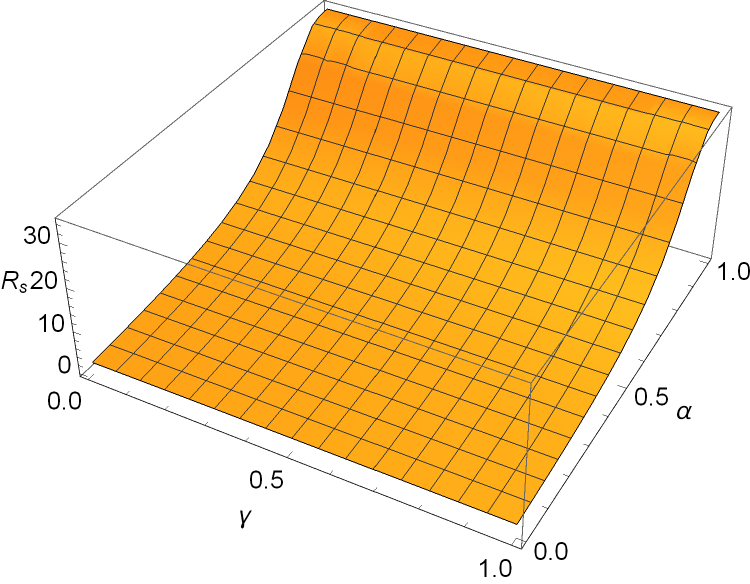}}\quad\quad
   {\centering{}\includegraphics[width=0.85\linewidth]{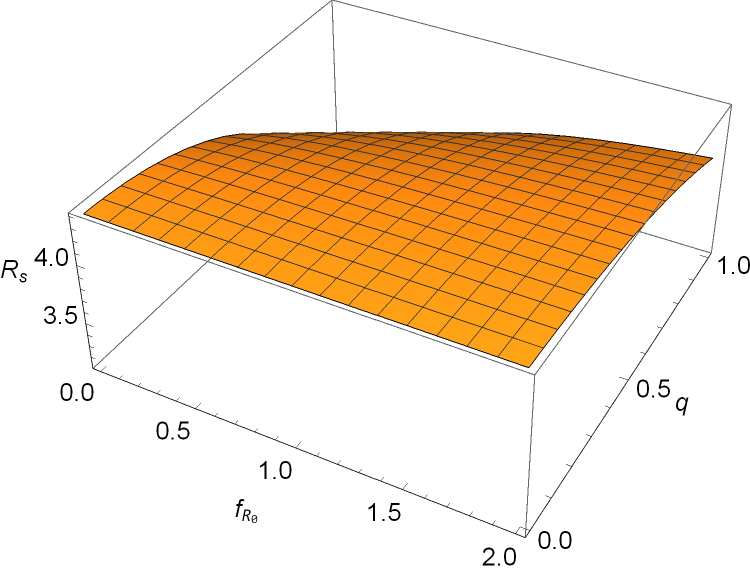}}
    \caption{Plot of shadow radius $R_{s}$ for different values of BH parameters. Here, we set $M=1$ and $R_0=-0.01$.}
    \label{figa22}
\end{figure}

To analyze the influence of the BH parameters on the shadow radius, we plot the shadow radius as a function of each parameter while holding the others fixed (see Fig. \ref{figa22}). In top panel, it   shows that an increase in $\gamma$ and $\alpha$ expands the shadow radius. However, in bottom panel, it  shows that increasing $q$ decreases the shadow radius. Notably, the CS parameter $\alpha$ has a stronger impact on shadow radius than other BH parameters.

 To represent the actual shadow of the BH as seen from an observer's perspective, we introduce celestial coordinates, $X$ and $Y$
\begin{equation}
X=\lim_{r_{\mathrm{o}}\rightarrow \infty }\left( -r_{\mathrm{o}}^{2}\sin
\theta _{\mathrm{o}}\frac{d\varphi }{dr}\right) ,\hspace{1cm} Y=\lim_{r_{\mathrm{o}}\rightarrow \infty }\left( r_{\mathrm{o}}^{2}\frac{%
d\theta }{dr}\right) .
\end{equation}%

For a static observer at large distance, i.e. at  $r_{\mathrm{o}}\rightarrow
\infty $ in the equatorial plane $\theta _{\mathrm{o}}=\pi /2$, the
celestial coordinates simplify to $X^{2}+Y^{2}=R_s^{2}.$ 

 Figure \ref{figa22} illustrates the shadow radius of the $F(R)$-ModMax theory with CS  in the celestial plane for different values of BH parameters. We observe that
 the shadow radius increases with increase in the parameter ($\gamma,\alpha, f_{R_0}$) however, the shadow radius decreases with increase in BH charge ($q$). 

\begin{figure}[ht!]
    \centering
    {\centering{}\includegraphics[width=0.45\linewidth]{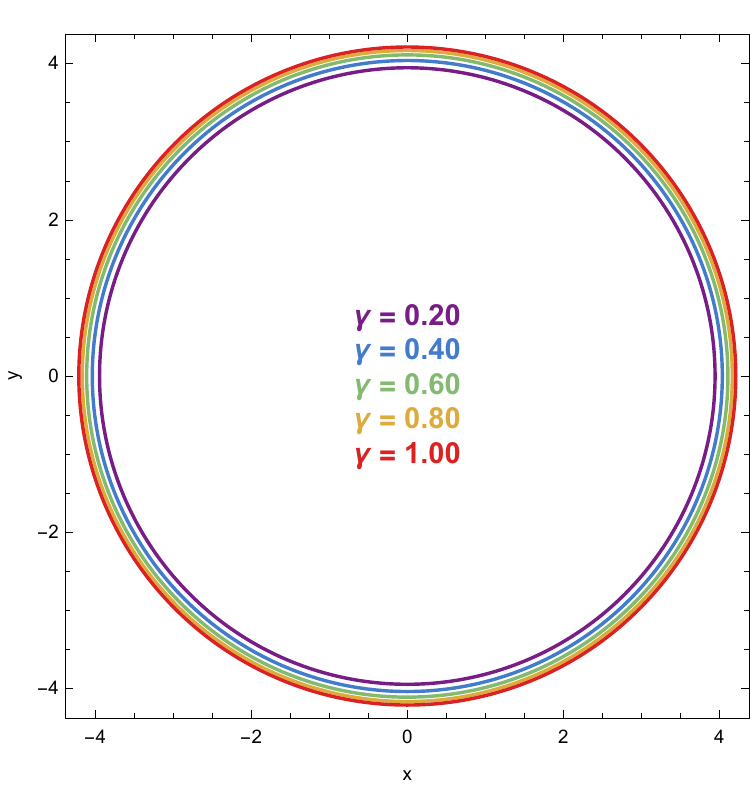}}\quad\quad
    {\centering{}\includegraphics[width=0.45\linewidth]{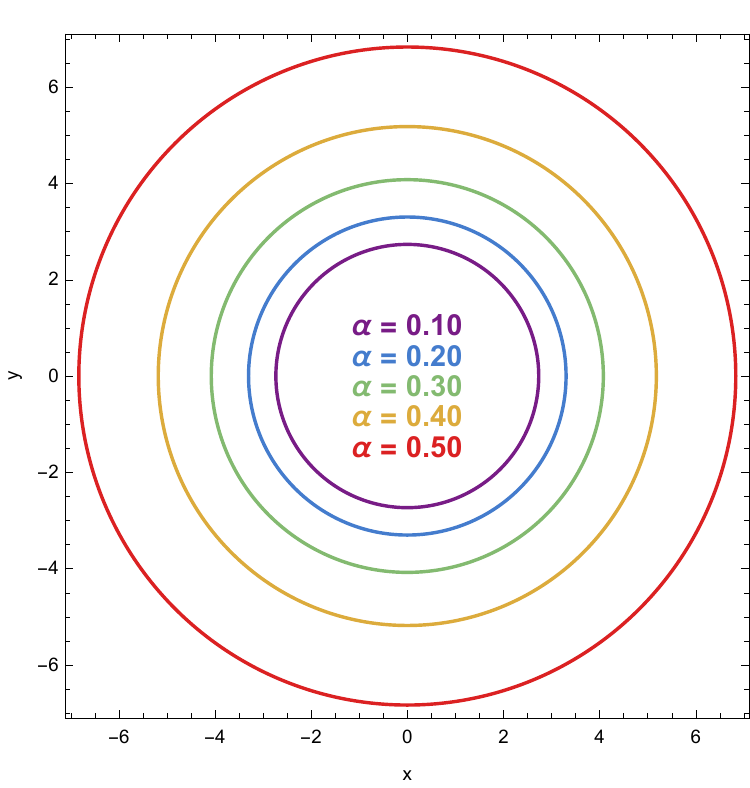}}\\
    {\centering{}\includegraphics[width=0.45\linewidth]{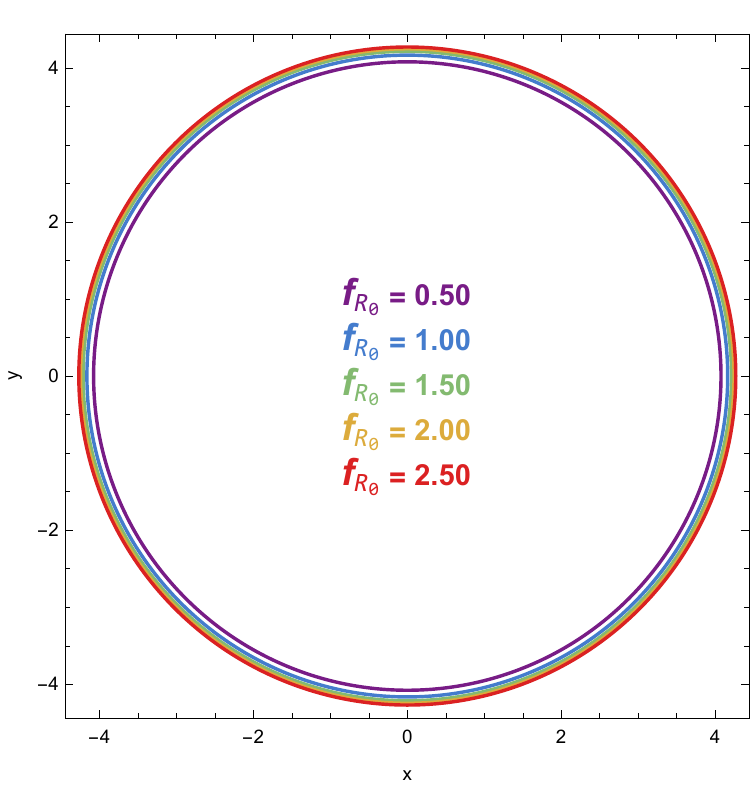}} {\centering{}\includegraphics[width=0.45\linewidth]{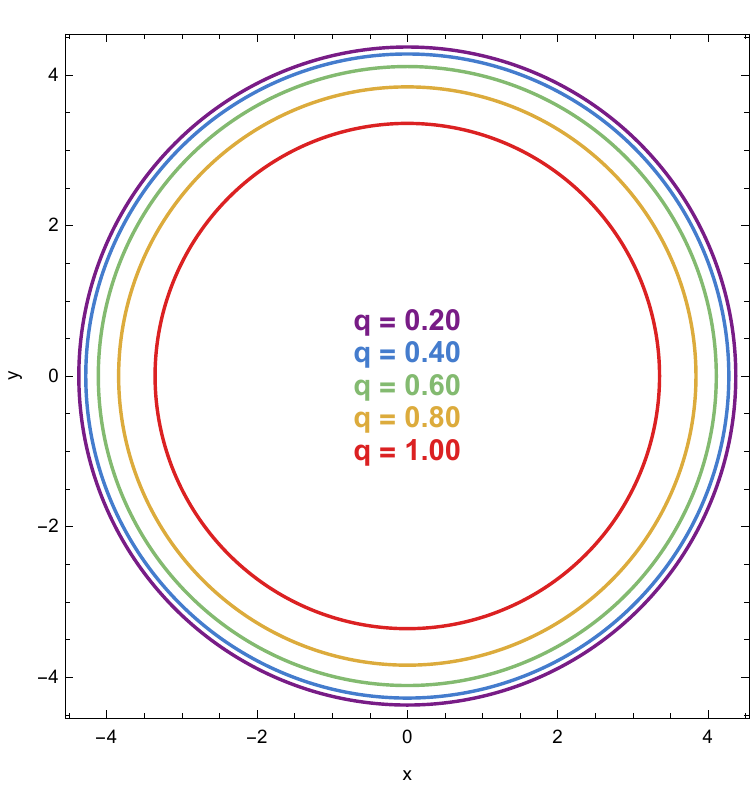}}
    \caption{BH shadow in the celestial plane for different values of BH parameters. Here, we set $m_0=1$ and $R_0=-0.01$.}
    \label{circle}
\end{figure}

\section{Conclusions}\label{sec5}

In this study, we obtained exact analytical solution to the gravitational field equations for the $F(R)$-ModMax theory in the presence of CS. This new BH solution is characterized by parameters including the BH mass 
$m_0$, charge 
$q$, the cosmological constant 
$R_0=4\,\Lambda$, the ModMax parameter 
$\gamma$, the 
$F(R)$  parameter $f_{R_0}$, and the CS parameter $\alpha$. Then, we studied the combined effect of the $F(R)$-ModMax theory and the CS on the BH shadow.

The analysis  explored how various physical parameters influence the BH’s shadow. We derived analytical solutions for both the photon sphere and shadow radius. Our results show that as the parameters ($\gamma,\alpha,f_{R_0}$) increase, so do the photon sphere and the shadow radius (Figure \ref{figa22}). In contrast, a higher charge parameter $q$ leads to a reduction in this radii. The shadow size of the $F(R)$-ModMax BH with CS is highly sensitive to the BH parameters (Fig. \ref{circle}). For large values of 
($\gamma,\alpha,f_{R_0}$)
 , the shadow expands significantly, whereas increasing 
$q$ causes it to shrink. Notably, the parameters ($\gamma,\alpha,f_{R_0}$) and the charge 
$q$ consistently exhibit opposing effects on physical quantities. Higher ($\gamma,\alpha,f_{R_0}$) tend to produce behavior more consistent with standard BHs.\\ {Our results indicate that the CS parameter ($\alpha$) leads to a significant and potentially distinguishable increase in the shadow radius. This provides a unique signature that could help in the interpretation of BH shadow measurements. For instance, an unexpectedly large shadow for a BH of a given mass could be explained by the presence of a CS rather than, or in addition to, other parameters.
 Table \ref{tableC1} provides a systematic comparison of photon sphere ($r_{ph}$) and shadow radii ($R_{s}$) across various theoretical frameworks. The results reveal a clear hierarchy of gravitational effects. The most significant finding is the dramatic enhancement of both $r_{ph}$ and $R_{sh}$ under the inclusion of CS, which increases their values compared to their counterparts in general relativity and other modified theories.
\begin{center}
\begin{tabular}{|c|c|c|}  \hline
Spacetime geometry & $r_{ph}$ & $R_s$ \\ \hline
$F(R)$-ModMax with CS & 3.52158 & 9.43396 \\ 
$F(R)$-ModMax  & 1.24066 & 2.26597 \\ 
ModMax  & 1.03261 & 2.02987 \\ 
Reissner-Nordstr\"{o}m  & $2.82288$ & $4.96791$ \\ 
Schwarzschild  & $3$ & $5.19615$ \\
 \hline
\end{tabular}
\captionof{table}{\footnotesize  A comparison of the radii $r_{ph}$ and $R_{s}$ for different spacetime geometries. Here, $\gamma=0.4$, $q=0.6$, $f_{R_0}=0.5$, $m_0=1$ and $R_0=-0.01$.} \label{tableC1}
\end{center}
In addition, as discussed in studies of dark matter spikes and other exotic matter configurations \cite{new5,new6,new7}, anisotropic distributions of matter can introduce characteristic deformations in the BH shadow. In our solution, the string cloud, while modeled here as a static and spherically symmetric background, provides a fundamental theoretical framework for such an anisotropy. A deviation from the shadow size predicted by the Schwarzschild solution could, therefore, be indicative of not only a cosmic string environment but also a signature of the underlying $F(R)$ gravity theory.} 

Finally, this research highlights several exciting directions for future exploration. First, calculating the quasinormal modes could reveal important details about the gravitational wave emissions from such BHs, opening possibilities for observational verification. Second,  study optical features such as energy emission rate and deflection of light  would offer a deep insight into recent astrophysical observation. Third, a fascinating topic will be examining the thermodynamic geometry of the $F(R)$-ModMax with CS as well as the geodesics structure. These avenues will be the focus of our upcoming work.

\end{document}